\documentclass[a4paper,fleqn]{cas-dc}

\usepackage[numbers]{natbib}
\def\tsc#1{\csdef{#1}{\textsc{\lowercase{#1}}\xspace}}
\tsc{WGM}
\tsc{QE}
\begin{document}
\let\WriteBookmarks\relax
\def\floatpagepagefraction{1}
\def\textpagefraction{.001}
\let\printorcid\relax % 可去掉页面下方的ORCID(s)

% Short title
% \shorttitle{<short title of the paper for running head>} 
\shorttitle{Design and optimization of a novel leaf-shape antenna for RF energy transfer}    

% Short author
% \shortauthors{<short author list for running head>}
\shortauthors{Zhengbao Yang et al.}

% Main title of the paper
\title[mode = title]{Design and optimization of a novel leaf-shape antenna for RF energy transfer
}

\author[1]{Junbin Zhong}%[style=chinese]
%\fnmark[1]

\author[1]{Mingtong Chen}%[role=Co-ordinator, suffix=Jr]
%\fnmark[2] 
%\ead{rishi@sayahna.org}
%\ead[URL]{www.sayahna.org}
%\credit{Data curation, Writing - Original draft preparation}

\author[1]{Zhengbao Yang}
\cormark[1] 
%\fnmark[3]
\ead{zbyang@hk.ust} 
\ead[URL]{https://yanglab.hkust.edu.hk/}

\address[1]{The Hong Kong University of Science and Technology
Hong Kong, SAR 999077, China}
% \address[2]{Sayahna Foundation, Jagathy, Trivandrum 695014, India}
% \address[3]{\TeX{} Users Group, Providence, MA, USA}

\cortext[1]{Corresponding author} 
%\cortext[2]{Principal corresponding author} 

% Here goes the abstract
\begin{abstract}
In this research, the design and optimization of a novel leaf-shaped antenna inspired by natural leaf structures for radio frequency energy transfer is presented. The objectives of this study are to develop a bio-inspired antenna, optimize its performance through impedance matching for the 915 MHz frequency band, and evaluate its efficiency in capturing RF energy. The design process involves selecting an appropriate leaf shape, modeling the antenna using AutoCAD and HFSS software, and fabricating a printed circuit board (PCB) prototype. Simulations and physical tests are conducted to optimize the antenna's performance, achieving an S11 parameter of nearly -20 dB at 915 MHz, indicating effective energy capture. Experimental results demonstrate the antenna's ability to power a device at distances up to 200 cm, with charging times reflecting its efficiency. The study concludes that the bio-inspired design of the proposed antenna improves RF energy transfer. Future work should focus on testing the antenna's penetration through concrete and developing a feedback system for autonomous alignment.

\end{abstract}

% Use if graphical abstract is present
%\begin{graphicalabstract}
%\includegraphics{}
%\end{graphicalabstract}

% Research highlights
% \begin{highlights}
% \item highlight-1
% \item highlight-2
% \item highlight-3
% \end{highlights}

% Keywords
% Each keyword is seperated by \sep

\maketitle

% Main text
\section{Introduction}

As one of the most widely used building materials in the world, concrete is renowned for its exceptional performance in aspects like compressive strength, durability and cost-effectiveness. These good attributes make concrete almost indispensable for most structures. However, despite the robust performance concrete shows, its integrity should not be taken for granted. Regular inspections of concrete structure are essential for monitoring the health of concrete structure and detecting potential problems such as cracks, corrosion and material degradation. These regular inspections help to ensure the longevity and safety of the concrete structure.

To monitor the long-term behavior of concrete structures, inserting small sensors inside or on the surface of the concrete can be regarded as one of the most promising research directions\cite{1}. After years of research, there are now more than fifty types of different sensors that are deployed for continued measurement of concrete properties like structural status, corrosion status and moisture state\cite{2}. Other novel types of optic fiber-based sensor for concrete monitoring utilizing technologies like fiber Bragg grating\cite{3} and Fabry - Pérot interferometer\cite{4}are also in rapid development.
In the research area of concrete embedded sensors, one crucial problem is the power supply for the sensor. Powering embedded sensors in concrete with batteries has several drawbacks. Batteries for the embedded sensor have limited lifespans, which can be as short as three months\cite{5}. This can be problematic for long-term monitoring in concrete structures. Replacing batteries in embedded sensors is almost impossible due to the obstruction of the concrete block. Battery will degrade rapidly under harsh environmental conditions like highly humid and alkaline environment commonly found inside concrete, further shortening the service life of battery. The size and weight of batteries is also a factor hindering the miniaturization of the sensor.

As described before, battery is not suitable for embedded sensors. The environment inside concrete also makes light, thermal, and mechanical energy sources inadequate or completely absent, thus unable to serve as the main power supply\cite{6}. To settle such problem, energy technologies like radio frequency energy harvesting (RFEH) and wireless power transfer can be an alternative solution for the embedded sensor. These two methods make use of the far field radio frequency (RF) to supply power. RFEH technology captures and converts ambient radio frequency signals into usable electrical power. It typically involves using antennas to capture RF signals from various sources such as television broadcasts and cellular networks. These signals are then rectified and converted into direct current (DC) electricity. WPT technology refers to the technology of transmitting electrical energy from a power source to a load without physical connections through electromagnetic fields. WPT involves several working principles, and can be classified according to the coupling method, namely inductive coupling, EM radiation and magnetic resonant coupling\cite{7}. 

The receiving side of radio frequency energy harvesting (RFEH) and radiative wireless power transfer (WPT) system typically consists of a receiving antenna, a matching network, a rectification circuit, and a power management unit\cite{8}. The combination of receiving antenna and the rectification circuit is often referred to rectenna. The radiation performance of receiving antenna significantly influences the RFEH or WPT system. The antenna is the key component of a rectenna, since it takes up the task of capturing RF signals, thereby influencing the overall performance of the RFEH or WPT system to a great extent. Nevertheless, the impedance matching between the rectifier circuit and the antenna also has an impact on optimal efficiency.

As the antenna part plays a significant role in the receiving side of the RFEH and WPT system, an appropriate design of the antenna is crucial. In this study a novel, bio-inspired design of receiving antenna for WPT purpose is proposed. The shape of the receiving antenna derives from leaves found in nature. The performance of the proposed antenna is evaluated by simulation software on computer and physical measurement of the manufactured sample in reality. After the measurement, methods to enhance the performance of the antenna like impendence matching is used to make the optimize the operating frequency band of the antenna and its performance within the operating frequency band.

Wireless power transfer (WPT) represents a transformative method that doesn't rely on traditional wired connections to deliver electrical energy. This technology provides a solution for powering devices in various distances and application scenarios. The WPT technology can be roughly divided into two main categories, near-field power transfer and far-field power transfer. Each category uses different mechanisms to achieve energy transfer\cite{9}. The near-field power transfer includes inductive power transfer (IPT) which transmits energy through magnetic field and capacitive power transfer (CPT) which transmits energy through electric field\cite{10}. Near-field power transfer is practical in short-distance scenarios and are usually used to charge consumer electronics such as smartphones. Far-field power transfer, which can be further divided into two categories of microwave power transfer (MPT) and laser power transfer (LPT), are designed for transferring energy at a longer distance. Among these technologies, microwave power transfer stands out as a promising far-field method, which uses microwave radiation to transfer power.

The working principle of microwave power transfer can be briefly described as follows. A DC source is fed to the transmitting side, and electrical power is converted into RF output power, which provides excitation for a transmitting antenna and then be transmitted in the form of intensified microwave beam. On the receiving side, a receiving antenna captures the transmitted microwave beam and transfers then to a rectifier, which converts the received RF power back into usable electrical energy. This process enables microwave power transfer to deliver energy over long distances, which can be as long as several kilometers. One of its key advantages is its adaptability to different environmental conditions, allowing it to operate effectively in environments where wired infrastructure is not feasible. Early experiments by Brown et al. demonstrated MPT’s potential, achieving 54\% efficiency at 2.446 GHz in the 1970s. Recent research topics for MPT include optimizing architecture of the transmitter, improving rectenna design for better RF to DC conversion, etc. The phased arrays integrated with high-efficiency GaN-based power amplifiers developed by Yamaguchi et al. achieved over 70\% efficiency at 5.8 GHz\cite{11}. 

Nature has been a rich resource library for many fields of engineering design for a long time. After the long evolution process measured in millions of years, animals and plants are both plentiful resources of inspiration for researchers who looking for surpassing the limits of conventional design towards more intricate and robust solutions. Reception of solar energy, which is almost the only way for plants to obtain energy necessary for survival, is crucial for plants as plants receive sunlight encompassing a wide frequency range of light but are only able to photosynthesize by utilizing certain wavelengths, mainly in blue and red vision bands due to the characteristics of chlorophyll pigments. Therefore, plants’ leaf has widely taken on a flat and broad shape throughout the evolution process. This shape is highly advantageous as it enables them to effectively capture light while minimizing self-shading\cite{12}. 

The geometric shapes of leaves are intricately linked to their ecological functions and developmental processes. To be specific, these shapes play a crucial role in optimizing light absorption for photosynthesis and facilitating the efficient conduction of water, both of which are essential for the plant's survival and growth\cite{13}. Therefore, some researchers believe that a plant can be regarded as a natural wireless system\cite{14}. Throughout the years there have been many studies on the topic of bioinspired leaf-shaped antenna, and remarkable progress in achieving wider bandwidth, better miniaturization and higher performance have been made in recent years. Fakharian and Rezaei introduced a compact planar monopole antenna based on a palmate leaf geometry, fabricated on an FR4 substrate, which exhibited ultra-wideband (UWB) characteristics from 3 to over 14 GHz with a stable omnidirectional H-plane radiation pattern and an average gain of 3 dBi, making it suitable for UWB applications\cite{15}. Oliveira et al. use polar transformation to generate a dissimilar four-leaf clover shaped antenna array, which enhances the antenna’s bandwidth to 81 MHz and maximum gain to 8.24 dBi\cite{16}. 

Çelik K and Kurt E reported their study on a novel rectenna design and implementation based on the shape of the Linden leaf shape, showing potential for energy harvesting application from 1.6 to 2.65 GHz frequency\cite{17}. Jothi Chitra et al. proposed a maple leaf-shaped microstrip patch antenna excited with 50$\omega$ feed line, resonating for 6.6 to 8.9 GHz band frequency, demonstrating good VSWR value and return loss enough for wireless application\cite{18}. Abolade et al. proposed a semi-Vine-leaf-shaped microstrip antenna with arc slits, enabling hexa-band operation at frequencies including 2.37 GHz, 3.06 GHz, and 4.88 GHz, tailored for ISM, WiMAX, and 5G mid-band applications, with a compact footprint of 30 × 12 mm\cite{19}. Cruz et al. presented a printed monopole antenna inspired by the Inga Marginata leaf, covering an ultra-wideband range from 340 MHz to over 8 GHz, whose frequency, gain, detection sensitivity and immunity is proven to meet the requirement for detecting partial discharge in high-voltage insulation systems, and is suitable for substation monitoring applications\cite{20}. Harish and Sudhir discussed the potential for their crop leaf-based antenna to be embedded in WSN node operating in ISM band\cite{21}, showing that leaf-shaped antenna holds potential for integration with emerging technologies like the Internet of Things. Together, these studies demonstrate good bandwidth, compact size, and versatile applications of leaf-shaped antennas driven by innovative bio-inspired designs. A review article makes a comparative analysis of several types of bio-inspired antenna, including leaf-shaped antenna, with traditional microstrip antenna, concluding that bio-inspired antennas outperform traditional microstrip antennas in key metrics like bandwidth, size and gain\cite{22}.

\section{Design Process}

\begin{figure}[h]
	\centering
		\includegraphics[scale=0.8]{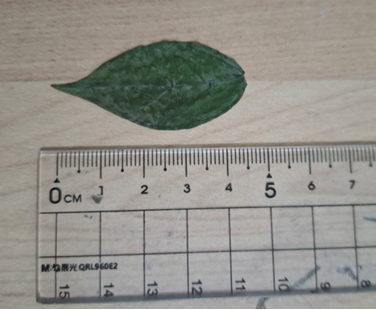}
	  \caption{The leaf selected to be modeled}\label{FIG:1}
\end{figure}

\begin{figure}[h]
	\centering
		\includegraphics[scale=0.6]{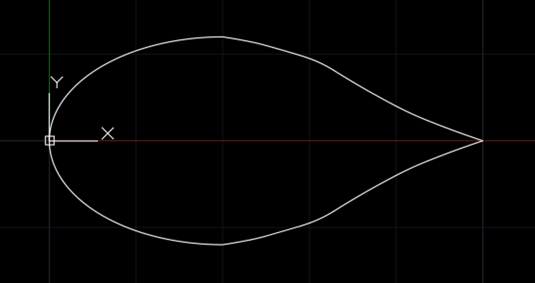}
	  \caption{The fitted pattern on AutoCAD}\label{FIG:2}
\end{figure}

The antenna designed in this study is based on a device developed by a doctoral student in the laboratory for RF wireless energy transmission, aimed at creating an efficient receiving antenna to capture the RF energy sent from the transmitter side. The device features a transmitter side that can send out radio wave when powered by a battery, and a receiving side is equipped with a rectifier module, an electrical appliance - a CC430 chip and a ceramic antenna, which can measure the temperature and transmit the collated data through the ceramic antenna. The major aim of this study is to develop a receiving antenna that can efficiently capture the RF energy transmitted by the transmitter side. What’s more, as the energy transmitter mainly transmit energy in the 915 MHZ band, it is necessary to apply impedance matching for the proposed antenna to achieve the best reception in the 915 MHz band.
When determining the antenna‘s shape, the doctoral student suggested drawing inspiration from fractal or naturally occurring structures. After evaluating various factors, I chose to model the antenna after the shape of a leaf. This decision was driven by several key reasons. First, leaves feature a large surface area and a complex, often fractal-like structure, which can improve the antenna’s ability to capture RF signals from multiple directions and across a broad range of frequencies. Second, leaves naturally evolved to efficiently absorb electromagnetic radiation in the form of light, a principle that may effectively extend to capturing RF waves, which are also electromagnetic in nature. These reasons make us believe that the leaf-inspired design a promising approach for enhancing the performance of the RF energy receiver.

One of the key steps in designing the leaf-shaped antenna is how to mimic the shape of the leaf on the PCB board. According to the past existing researches, when it comes to this problem, the approaches used by researchers can be broadly divided into two groups. One the one hand, some researchers utilize specific chemicals along with image recognition technology to process the leaf. With this method, the real shape of a leaf and its complex vein network can be reproduced on a PCB board . One the other hand, researchers mimic the outer contour of a leaf blade by combining simple geometric shapes. Compared with the latter method, the former method allows for a more faithful reproduction of the specific shape details of the leaf, but at the expense of a much more complex process of handling the leaf. What’s more, the complex structures produced by the first method can make the simulation of the design extremely complicated and difficult. Taking these factors into consideration, I decided to use the latter method in imitating the leaf. 

During the period of selecting appropriate leaf to model, I evaluated several potential leaf candidates, comparing their size and shape, as the previous antenna is a patch antenna on a PCB board with a length and width of about 10cm*8cm, and I think the new antenna design should be at most equal in size to the original design. As a result, this requirement excludes many oversized leaves. Moreover, the shape is another important criterion, as shapes of some leaves substituted for features such as serrations are too irregular for shape characterization with simple geometric shapes. Finally, a leaf with proper size and shape is selected and is shown below.

After the selection of the leaf, the next step is to characterize the shape of the leaf on software. Here I used the AutoCAD developed by Autodesk. As mentioned above, the guiding principle of the characterization is using combination of simple geometric patterns and lines to fit the shape of leaf. Using the petiole as the starting reference, the first half of the leaf is found to be fitted well with half an ellipse with appropriate major axis and minor axis, which are measured from the leaf. The second half of the leaf is fitted using a series of points taken at different places along the edge of the leaf, which are then connected with a Bézier curve. Different combinations of sampling points were tested to achieve the best fit. The final graph after optimization is shown in the figure. To efficientlly use the space in  a limited size, the shape is rotated 45° counterclockwise and mirrored.

After the creation of the leaf shape is completed, the next step is to import the shape to the PCB design software. Here I use the EDA software by JLCPCB. The dxf file exported from AutoCAD is loaded into the EDA software. Then is time for PCB routing and placing surface mount device’s pad, which are reserved for impedance matching of the antenna in the future. Figure 3 shows the PCB design of the antenna. This antenna design shares some similarities with the traditional bowtie antenna, while having some distinct differences, too. Initially invented by Oliver Lodge in 1898, bowtie antenna is a broadband antenna, known for its simple and symmetrical design. This type of antenna usually features two triangular or trapezoidal elements flared from a central feed point. This geometry brings wide impedance bandwidth and stable radiation patterns, making it suitable for applications such as ultra-wide band communications . Compared with bowtie antenna, the proposed antenna has a pair of antenna elements which are configured symmetrically, and the feedline are located in the center of the antenna, similar to traditional bowtie antenna. But instead of being triangular, the antenna elements of the proposed antenna are leaf shaped. Moreover, instead of relying on the geometric configuration to realize impedance matching, which is the common practice of bowtie antenna design, the proposed antenna leaves space for matching components like inductors or capacitors, offering a more flexible impedance matching approach. The feedline stretches out from where the leaf connects to the petiole, mimicking real leaf, and then connects to a SMA connector, which can be further connected to various RF components or test equipment for measurement and system integration.

\begin{figure}[h]
	\centering
		\includegraphics[scale=0.9]{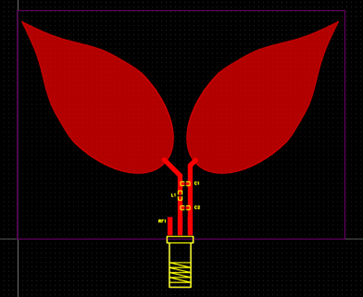}
	  \caption{ PCB design of the proposed antenna}\label{FIG:3}
\end{figure}

\section{Software simulation and physical Testing of optimization}

After completing the design work for the antenna, the next step is to run simulation of this antenna on computer to make a preliminary observation of the antenna performance. For the electromagnetic simulation analysis, the HFSS developed by ANSYS Inc. is used. Importing the gerber file from EDA software to HFSS, the structure of the antenna is restored in the HFSS software. Lumped RLC boundaries are implemented to simulate the inductor and capacitor to be used. A lumped port is set on the lower part of the SMA connector pads to simulate what the SMA connector receives from the antenna.

At first the thickness of the copper layer was taken into consideration. However, combined with the non-single simple geometric figure of the antenna elements, this makes the finite element analysis (FEA) process much more complex and lengthier, requiring more than 10 minutes to complete one simulation. Therefore, my mentor suggested me to treat the copper layer as a perfect E, which means the copper layer is treated as a perfect electrical conductor and can greatly speed up the simulation process. Whether a conductor can be treated as perfect E or not mainly depends on a characteristic namely skin depth, which represents the depth at which the current density decays to 1/e (about 37\%) of its surface value, is calculated as follows: 

$$
\delta=\sqrt{\frac{\rho}{\pi f \mu_0 \mu_r } } 
$$

After setting all the boundaries and excitations, the simulation process began. During the simulation, the S11 parameter is used for evaluating the performance of the antenna. The S11 parameter is a member of the S-parameter family. The S-parameters describe the input-output relationship between ports (or terminals) in an electrical system. Suppose there are 2 ports namely Port 1 and Port 2, then S12 represents the power transferred from Port 2 to Port 1 and S21 represents the power transferred from Port 1 to Port 2. In general, SNM represents the power transferred from Port M to Port N in a multi-port network. Back to S11 parameter, as its name suggests, it represents how much power is reflected from the antenna, therefore it is also known as reflection coefficient. The S11 parameter in decibels can be calculated as follows:

$$
S_{11} =10\log_{10}{(\frac{P_r}{P_i})}
$$

In the context of an antenna designed to receive RF energy, the S11 parameter directly indicates the antenna’s ability to capture incoming RF energy. A low S11 parameter reflects low reflected power, meaning most incident energy is coupled into the antenna and delivered to the rectifier circuit, which leads to a higher energy efficiency. Conversely, a high S11 indicates significant power reflection, resulting in poor energy capture and reduced system performance. Typically, an S11 value of no larger than -10 dB is considered acceptable, indicating that less than 10\% of the incident power is reflected back due to impedance mismatch.

After completion of modeling the PCB antenna in HFSS, simulation and optimization begin. This process can be briefly summarized into the following steps.First, the antenna without any impedance matching components is simulated to observe the S11 parameter. Unsurprisingly, the S11 parameter at 915 MHz is close to zero. This is reasonable as it is a raw design without any impedance matching. Then is the time for matching the impedance of the antenna to the operating band. Impedance matching is a fundamental concept in RF and antenna design that involves adjusting the impedance of a load, in this study is the antenna, to match the impedance of the source or transmission line, which is 50 ohms. The Smith Chart is used during impedance matching. The Smith Chart is a useful graphical tool in RF engineering for analyzing and designing impedance matching networks. By representing normalized impedance in its circular coordinate system, the Smith Chart can visualize the complex impedance of a load in relation to a transmission line's characteristic impedance.Second, I read the complex impedance corresponding to the 915MHz point in the Smith Chart of the simulation result. Then the complex impedance is input into the Smith Chart software developed by Prof. Fritz Dellsperger. The initial state of the current antenna will be plotted at the periphery of the circle.Third, try different matching components. Different components will move the point in the coordinate system in different ways, depending on whether they are connected in series or in parallel, as well as their types of component and specific values. Generally, series components move the impedance along constant-resistance circles, parallel components shift the admittance along constant-conductance circles. The overall goal is to move the initial point to the center point, which marks the completion of impedance matching. Depending on the specific starting location, usually more than one matching component needs to be used. A typical example of impedance matching process is shown in Figure 5.

\begin{figure}[h]
	\centering
		\includegraphics[scale=0.4]{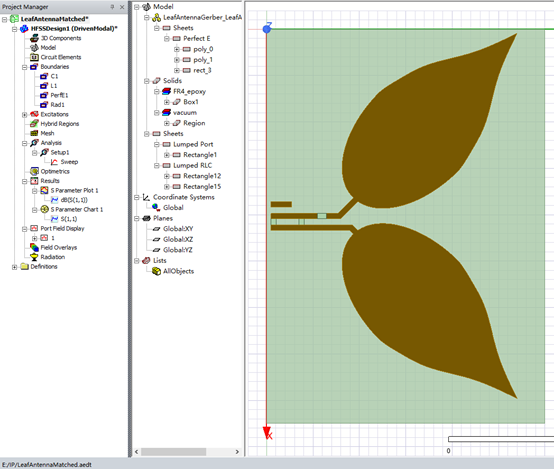}
	  \caption{Modeling the antenna in HFSS}\label{FIG:4}
\end{figure}

\begin{figure}[h]
	\centering
		\includegraphics[scale=0.5]{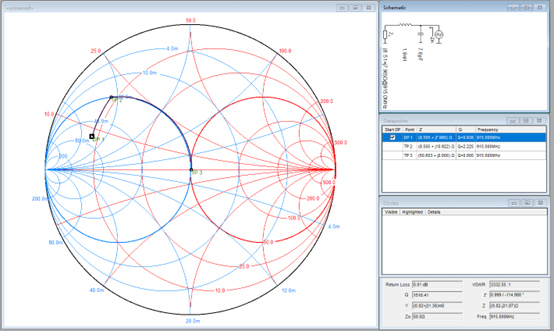}
	  \caption{Performing impedance matching with Smith}\label{FIG:5}
\end{figure}

\begin{figure}[h]
	\centering
		\includegraphics[scale=0.5]{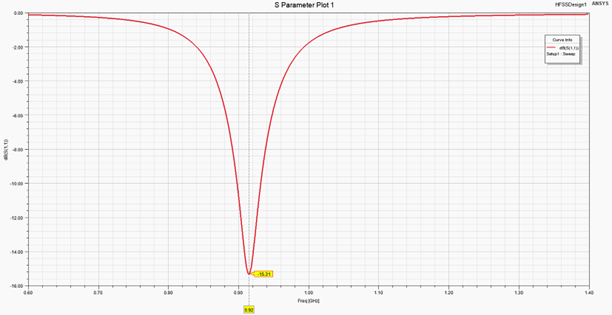}
	  \caption{S11 curve of the optimized impedance matching network}\label{FIG:6}
\end{figure}

\begin{figure}[h]
	\centering
		\includegraphics[scale=0.6]{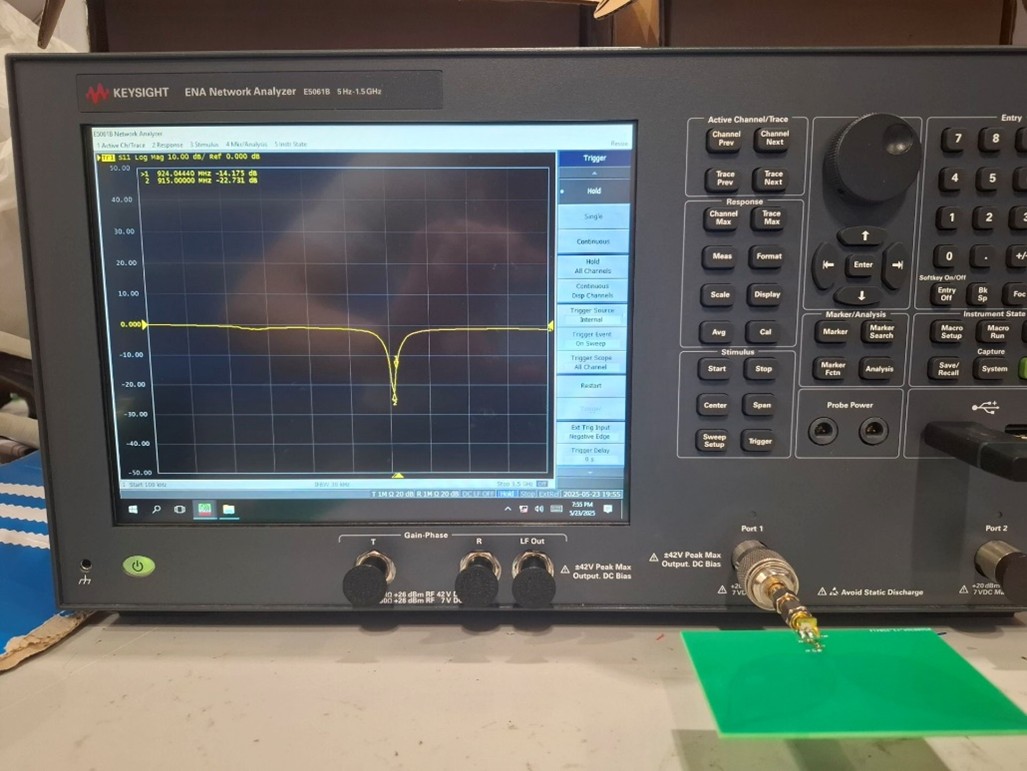}
	  \caption{Using ENA to analyzer antenna’s performance}\label{FIG:7}
\end{figure}

Fourth, upon moving the initial point to the center, the software will give out schematic of the impedance matching network, which contains what components are used, as well as their connecting sequences and specific values. Then place simulated matching components in HFSS according to the schematic and run the simulation again. This time the result is much better than the first simulation, and an obvious dip appears in the S11 diagram from simulation. However, the dip deviates a little bit from 915MHz, and the S11 parameter at 915 MHz still has some distance from being satisfactory. This is probably due to some difference between the HFSS and Smith software in the simulation process. To settle this problem, started from the original schematic, I tried several combinations of values of components around the original schematic’s component value. Finally, at a specific combination of matching components, the simulation result happens to be satisfactory – the dip of the S11 parameter curve points directly to 915MHz. The value of this set of components is recorded and will be used as the reference in implementing impedance matching on the physical antenna fabricated later.

Upon the completion of simulation, the next step is to manufacture the antenna and measure the performance of the physical antenna.
After the fabricated antenna is delivered, corresponding components are soldered onto the board according to the schematic from the simulation result. Then the performance of the physical antenna is tested via an ENA analyzer. The ENA analyzer is a versatile instrument widely used in RF and microwave engineering to measure the performance of devices like antennas, filters, and amplifiers, which is capable of analyzing scattering parameters (S-parameters), such as S11 and S21 to evaluate impedance, gain, and transmission characteristics.

However, this time the similar phenomenon of shifting in the best operating frequency occurred again, and the matching effect was significantly poorer. With the aid of mentor, I summarized the possible reasons for such discrepancy. The first reason is the idealized component models. In HFSS simulation I used RLC boundaries to represent SMT capacitors and inductors, which may have some variation in physical characteristics, ignoring effects like parasitic effects present in real components. The second reason is the discrete component value limitation. In HFSS I can use components with any value I want. However, in reality components are limited to specific, discrete values. For instance, if I need an inductor with 7.0nH, but only 6.8nH or 7.5nH options are available, then I have to resort to one that has the closest value. The third reason is the manufacturing tolerance. In reality components have tolerances, which can potentially make the actual value of the component deviate from its calibrated value. Nevertheless, the components’ configuration and values for impedance matching given by the simulation software are correct in the general direction, and the following work for me is to try different combination of real-world components around the given value and see if they perform well (Fortunately, the number of antennas I ordered is sufficient support such exhaustion). Finally, a combination is found to work well in physical antenna. Figure 8 gives out the schematic of the impedance matching network. Actually, the measurement output of the ENA analyzer indicates the actual performance of the antenna is even better than in the simulation. As is shown in Figure 9, the bottom of the dip of the S11 curve reaches near -40 dB, although the dip deviated a little and is not pointing exactly at 915 MHz, the S11 parameter at 915 MHz is still near -20 dB, which is acceptable enough for this study.

\begin{figure}[h]
	\centering
		\includegraphics[scale=1]{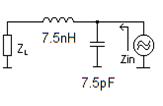}
	  \caption{ L-type matching network}\label{FIG:8}
\end{figure}

\begin{figure}[h]
	\centering
		\includegraphics[scale=0.7]{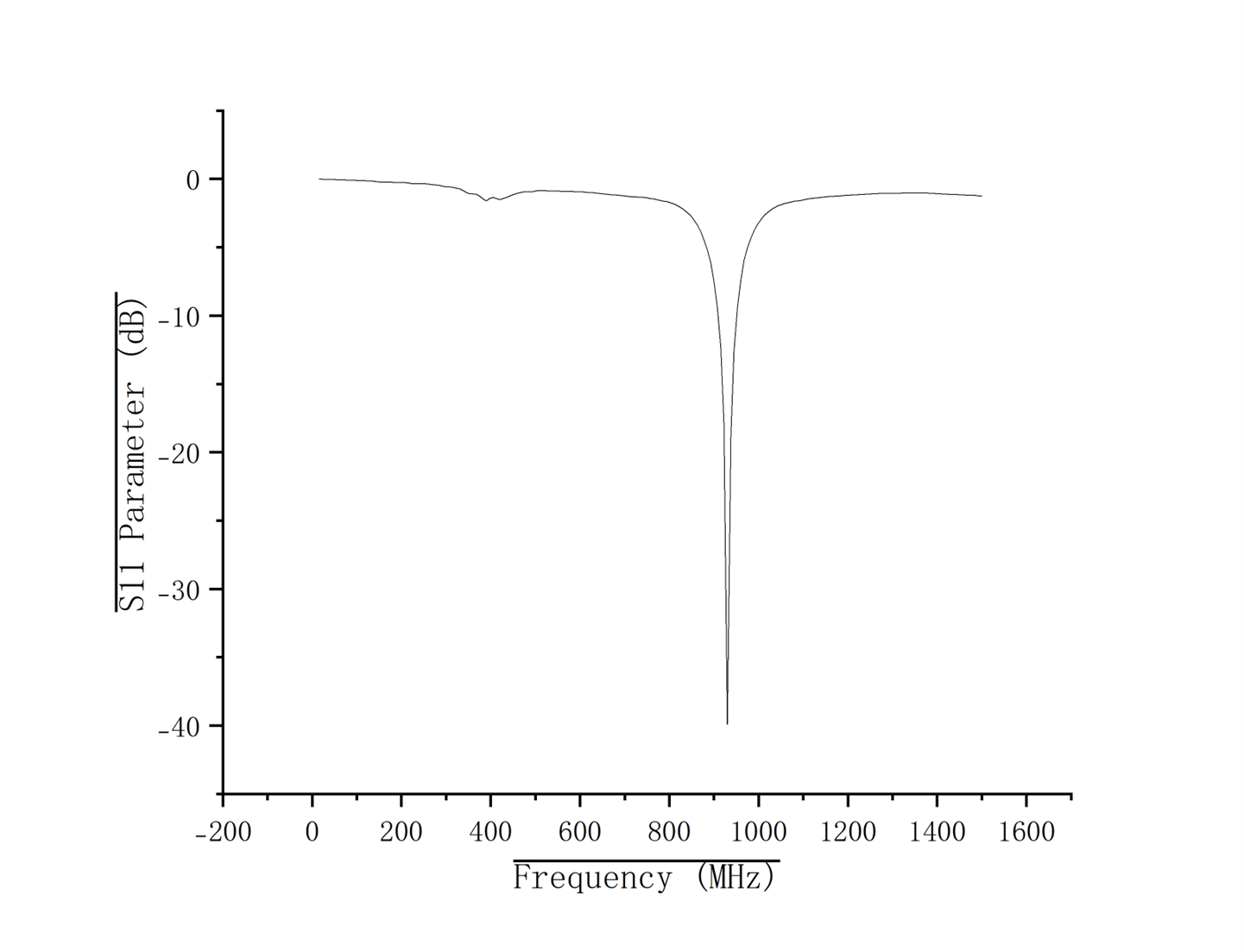}
	  \caption{S11 parameter measured by ENA analyzer}\label{FIG:9}
\end{figure}

\begin{figure}[h]
	\centering
		\includegraphics[scale=0.8]{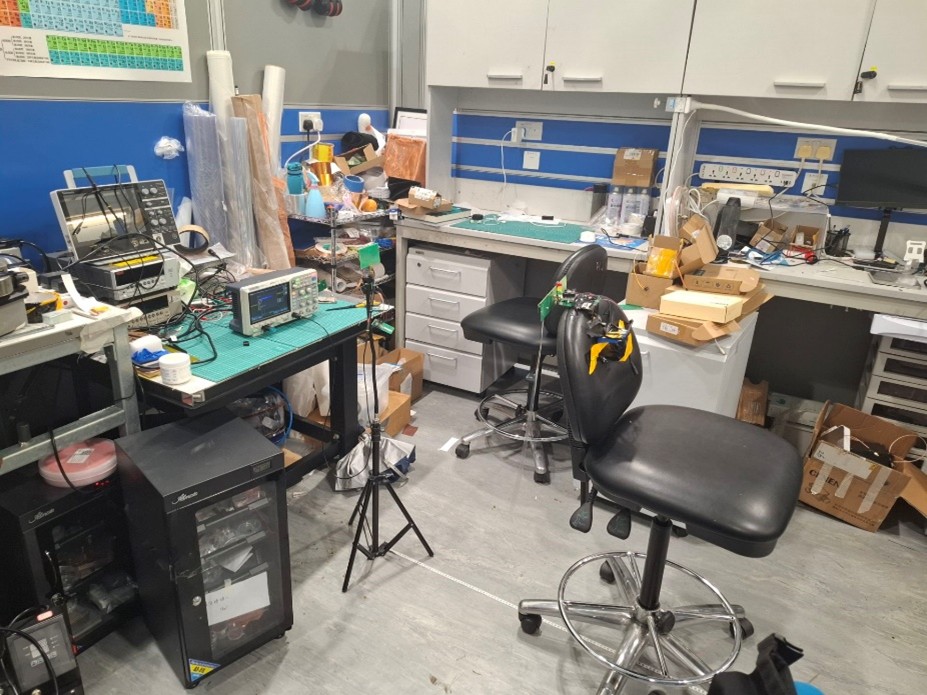}
	  \caption{Experiment Setup}\label{FIG:10}
\end{figure}

To test whether the proposed antenna can work with the device it base on, an oscilloscope is used. The receiver side of the device is comprised of a rectifier module and a CC430 chip, and the latter one is the power user. According to the senior student who designs and manufactures the device, the CC430 chip needs 4 V to start operation, which can be observed on the oscilloscope with a drop in voltage. Each voltage drop indicates the chip operates and sends out message for one time through the ceramic antenna beside the chip. The probe of the oscilloscope is connected to two pins of the electrolytic capacitor installed on the rectifier module.

During the experiment, the receiving side connected to oscilloscope with the proposed antenna is fixed on one side and the transmitting side is moving to record data at different distance. In the experiment the oscilloscope measures the potential difference between the two pins of the connected electrolytic capacitor. The time taken for the electrolytic capacitor to charge to 4 volts for the first time at different distances reflects the amount of energy received by the antenna: the shorter it takes to charge to 4 V, the more energy the antenna catches. The two figures below show how distance obviously affects the charge time.The experiment measured the charging time in the range of 50 cm to 200 cm, with each interval being 25 cm. The sorted charging times are shown in the figure follows.

\begin{figure}[h]
	\centering
		\includegraphics[scale=0.8]{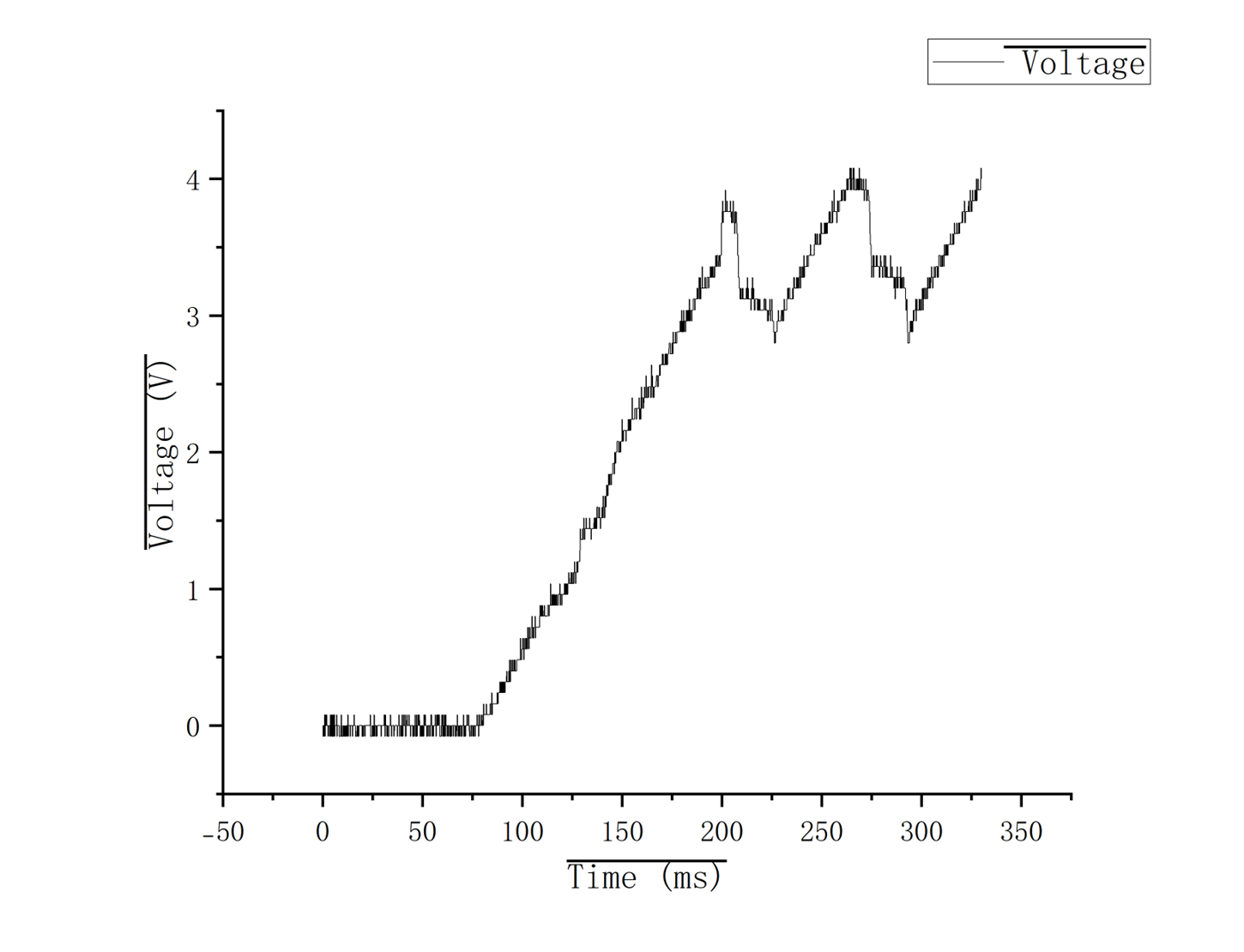}
	  \caption{charging time at 50 cm distance}\label{FIG:11}
\end{figure}

\begin{figure}[h]
	\centering
		\includegraphics[scale=0.6]{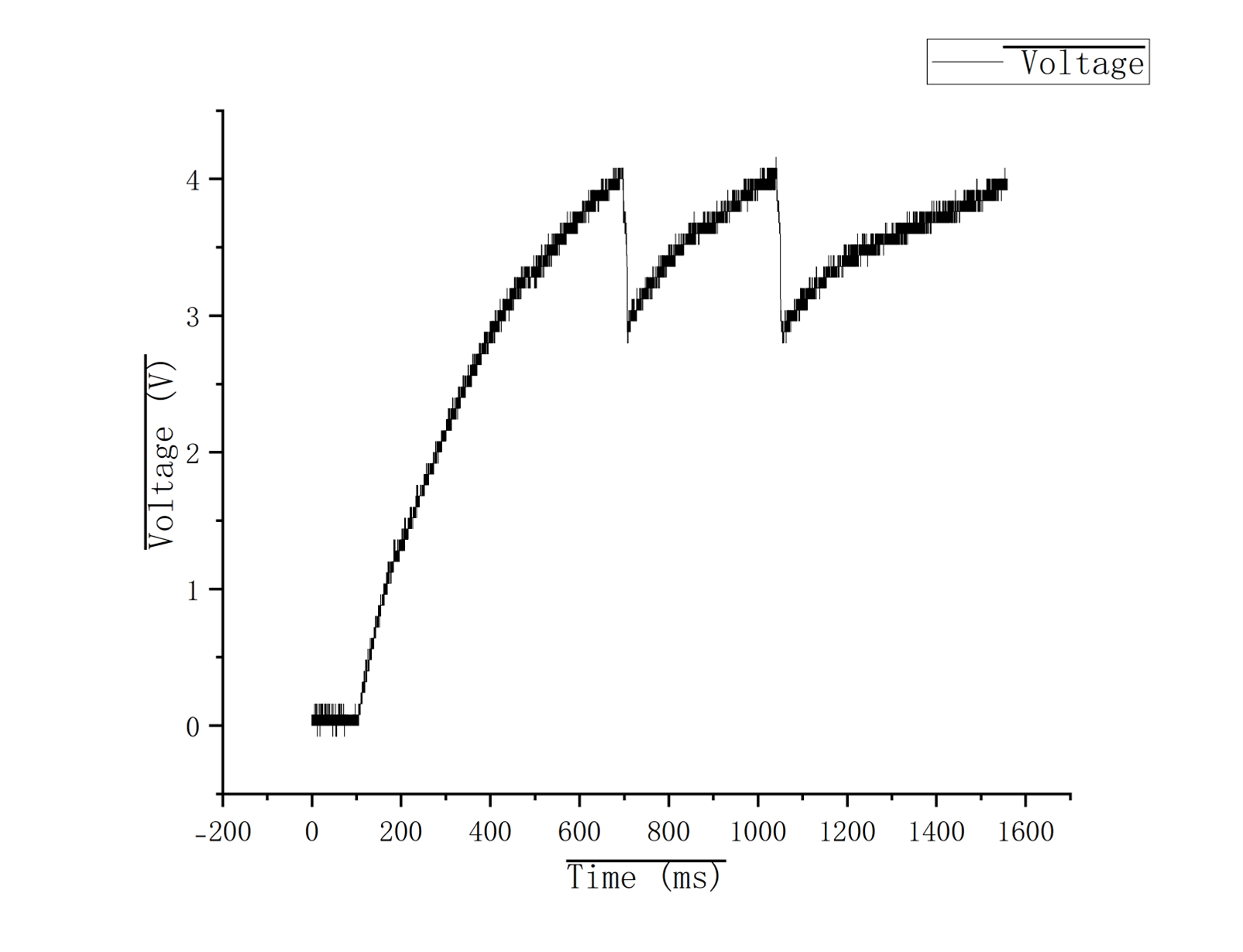}
	  \caption{charging time at 200 cm distance}\label{FIG:12}
\end{figure}

\begin{figure}[h]
	\centering
		\includegraphics[scale=0.6]{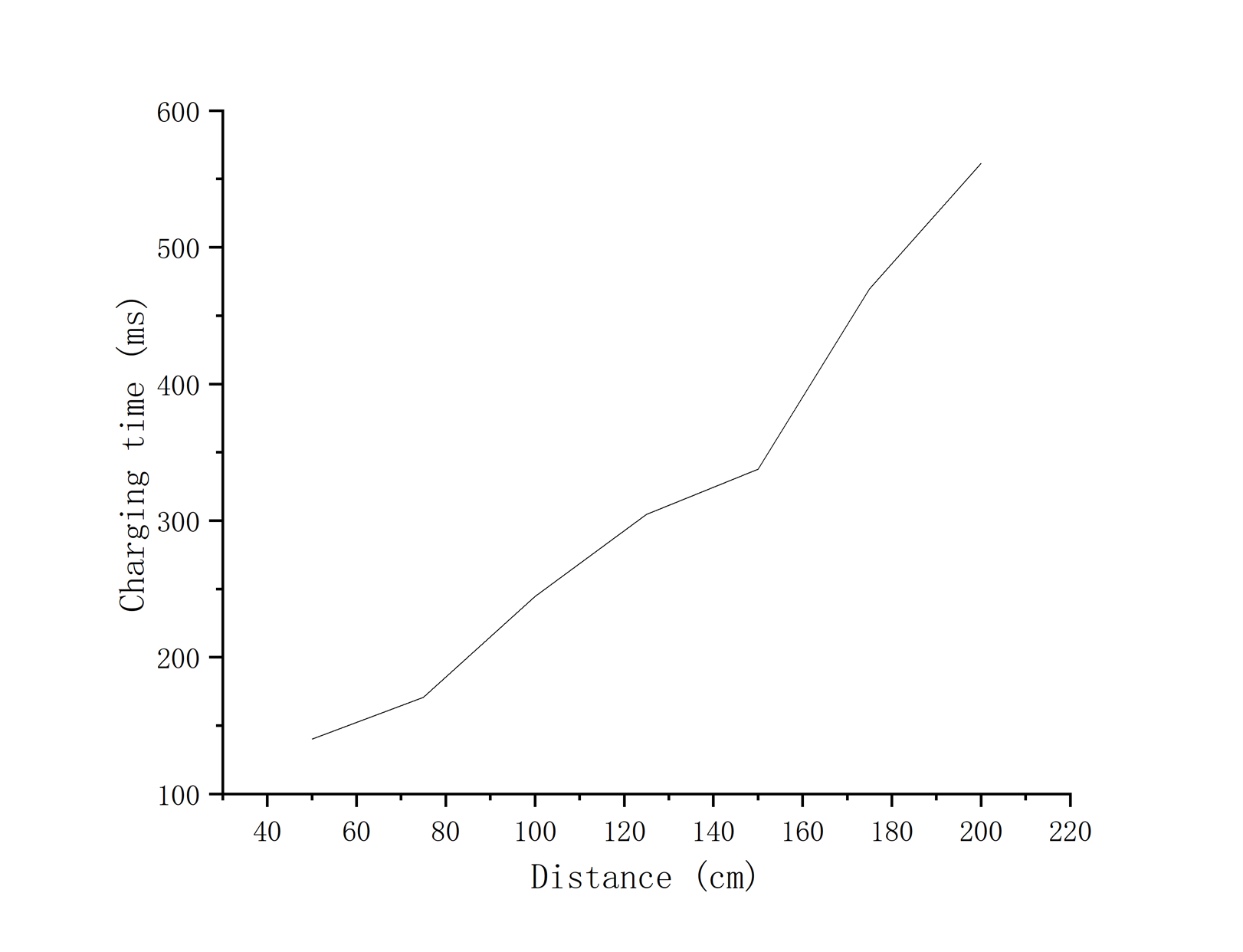}
	  \caption{summarized charging time}\label{FIG:13}
\end{figure}

\section{Conclusion}

This study briefly reviews past researches on related topics like leaf antenna and RF energy transfer at the beginning. Then the design and test process of a novel PCB antenna inspired by natural leaf and bowtie antenna is presented. After optimization methods like impedance matching the performance of the proposed antenna is believed to be at an acceptable level. However, further tests on the topic like penetration ability through concrete of the antenna should be carried out. Initially this study also intends to design a system with feedback mechanism to control a motor driving the transmitter to autonomously turning to the direction which the antenna can receive maximal RF energy. However, due to various technical difficulties and some unknown issues with the host computer software, which is essential for developing the system, the efforts to develop this system eventually ended in failure. In the future if condition permits, further studies can be carried out in designing such a system.

% Numbered list
% Use the style of numbering in square brackets.
% If nothing is used, default style will be taken.
%\begin{enumerate}[a)]
%\item 
%\item 
%\item 
%\end{enumerate}  

% Unnumbered list
%\begin{itemize}
%\item 
%\item 
%\item 
%\end{itemize}  

% Description list
%\begin{description}
%\item[]
%\item[] 
%\item[] 
%\end{description}  

%Figure
% \begin{figure}[h]
% 	\centering
% 		\includegraphics[scale=1]{elsevier-cas-double_column/wang_ruihua/f1.png}
% 	  \caption{Cardio axis}\label{fig:2}
% \end{figure}

% \begin{table}[h]
% \caption{Comparison of mechanical properties}\label{tbl1}
% \begin{tabular*}{\tblwidth}{@{}LL@{}}
% \toprule
%  Against Copper Film Test & Against Human Skin Test \\ % Table header row
% \midrule
%  Wet (10$\Omega$) & Wet (550$\Omega$) \\
%  Microneedle (1$\Omega$) & Microneedle (600$\Omega$) \\
%  Dry (0.1$\Omega$) & Dry (700$\Omega$) \\
% \bottomrule
% \end{tabular*}
% \end{table}

% Uncomment and use as the case may be
%\begin{theorem} 
%\end{theorem}

% Uncomment and use as the case may be
%\begin{lemma} 
%\end{lemma}

%% The Appendices part is started with the command \appendix;
%% appendix sections are then done as normal sections
%% \appendix

% To print the credit authorship contribution details
% \printcredits

%% Loading bibliography style file
%\bibliographystyle{model1-num-names}
\bibliographystyle{cas-model2-names}

% Loading bibliography database
\bibliography{cas-refs}

% Biography
% \bio{}
% % Here goes the biography details.
% \endbio

% \bio{pic1}
% % Here goes the biography details.
% \endbio

\end{document}